\documentclass[a4paper]{article}

\usepackage{INTERSPEECH2020}

\title{Improving EEG based continuous speech recognition using GAN}
\name{Gautam Krishna, Co Tran, Mason Carnahan, Ahmed H Tewfik}
\address{
  Brain Machine Interface Lab, The University of Texas at Austin
  }
\email{}

\begin{document}

\maketitle
\begin{abstract}
In this paper we demonstrate that it is possible to generate more meaningful electroencephalography (EEG) features from raw EEG features using generative adversarial networks (GAN) to improve the performance of EEG based continuous speech recognition systems. We improve the results demonstrated by authors in \cite{krishna2019improving} using their data sets for for some of the test time experiments and for other cases our results were comparable with theirs. Our proposed approach can be implemented without using any additional sensor information, whereas in \cite{krishna2019improving} authors used additional features like acoustic or articulatory information to improve the performance of EEG based continuous speech recognition systems. 
\end{abstract}

\section{Introduction}

Recently researchers have started exploring the possibility of synthesizing speech and text from neural signals. In \cite{anumanchipalli2019speech,angrick2019speech} authors demonstrated synthesizing intelligible speech from electrocorticography (ECoG) neural signals. Performing speech synthesis and speech recognition using neural signals might help people with speaking disabilities and difficulties to communicate with virtual personal assistants like Alexa, Bixby, Siri etc thereby improving technology accessibility and at the same time it will allow them to have normal conversation with their loved ones as well. In \cite{brumberg2018brain} authors proposed a brain–computer interfaces (BCIs) system that control a cursor to select letters one-by-one to spell out words but users can transmit only upto 10 words per min using their system, a rate slower than the average of 150 words per minute of natural speech whereas a continuous neural signal based speech recognition system would be capable of producing output at 150 words per minute. Electrocorticography (ECoG) is an invasive way of measuring electrical activity of brain where a brain surgery is performed to implant the ECoG electrodes. On the other hand  electroencephalography (EEG) is a non invasive way of measuring electrical activity of human brain. The EEG sensors are placed on the scalp of a subject to obtain the EEG recordings. Like ECoG, EEG also offer high temporal resolution even though the spatial resolution and signal to noise ratio (SNR) offered are lower compared to ECoG. Since EEG is a non invasive approach it is more safer and easier to deploy and study compared to ECoG. In \cite{krishna2020synthesis} authors provided preliminary results for synthesizing speech from EEG signals. In \cite{krishna20,krishna2019speech,krishna2019state} authors demonstrated continuous and isolated speech recognition using EEG signals for a limited English vocabulary in presence and absence of background noise. 

The results described by authors in \cite{krishna20,krishna2019state} demonstrate that continuous speech recognition using EEG is extremely challenging mainly due to the poor SNR offered by EEG signals. In a recent work described in \cite{krishna2019improving} authors proposed various techniques to improve the performance of EEG based continuous speech recognition systems. They show that by using an external language model and by adding deep layers in speech recognition encoder with their weights initialized with weights derived from an EEG to acoustic + articulatory regression model, will help in improving the recognition test time results. Even though their proposed method improved the results described in \cite{krishna20,krishna2019state}, additional sensors are needed to record acoustic or articulatory features to implement their method. 

In \cite{goodfellow2014generative} authors introduced the concept of generative adversarial networks (GAN) where two networks namely the generator model and the discriminator model which are trained simultaneously. The generator model learns to generate data from a latent space and the discriminator model evaluates whether the data generated by the generator is fake or is from true data distribution. The training objective of the generator is to fool the discriminator. In this paper we show that the concept of GAN can be used to generate more meaningful EEG features from raw EEG features to improve the performance of EEG based continuous speech recognition systems. 

We improve the results demonstrated by authors in \cite{krishna2019improving} using their data sets for for some of the test time experiments and for other cases our results were comparable with theirs. Our proposed approach can be implemented without using any additional sensor information, whereas in \cite{krishna2019improving} authors used additional features like acoustic or articulatory information to improve the performance of EEG based continuous speech recognition systems.

\section{Generative Adversarial Network Model}

The main motivation behind this idea is in the case of GAN the loss function is learned where as in case of an automatic speech recognition (ASR) model a fixed loss function like cross entropy \cite{chorowski2015attention} or connectionist temporal classification (CTC) loss \cite{graves2014towards} is used. However GAN models are extremely difficult to train. 

Our generator model, as shown in Figure 1, is very similar to the encoder part of the CTC ASR model described by authors in \cite{krishna2019improving}. However we initialize the gated recurrent unit (GRU) \cite{chung2014empirical} layers with random weights. After the temporal convolutional network (TCN) \cite{bai2018empirical} layer an average pooling layer is used.
The average pooling layer calculates the average value of all the time step outputs of the TCN layer. The average pool layer output is passed to dense layer with two hidden units and softmax activation function to produce fake label tokens. The label tokens were one hot vector encoded.  

The discriminator model, as described in Figure 2, consists of
a dense layer with 64 hidden units with rectified linear unit (ReLU) \cite{xu2015empirical} activation function and a single layer GRU with 128 hidden units connected in parallel. At each training step a pair of inputs are fed into the discriminator. The discriminator takes (real EEG features, fake label) and (real EEG features, real label) pairs.  The last time step output of the GRU layer and dense layer output are concatenated and then fed into a dense layer with sigmoid activation function. The  real EEG features are fed into GRU layer as input and fake, real label vectors are fed into the dense layer as input. 

In order to define the loss functions for both our generator and discriminator model let us first define few terms. Let $P_{s_f}$ be the sigmoid output of the discriminator for (real EEG features, fake label) input pair and let $P_{s_e}$ be the sigmoid output of the discriminator for (real EEG features, real label) input pair during training time. Then we can define the loss function of generator as $-\log (P_{s_f})$ and loss function of discriminator as $-\log (P_{s_e}) - \log(1-P_{s_f})$. 

The intuition here is since GAN learns the loss function also during training, the generator which is similar to the encoder of the CTC ASR model in \cite{krishna2019improving} will learn the most accurate EEG to label or text mapping. Especially the TCN layer in generator will learn the fine representations of input EEG features which are easily mapped to labels. 
During test time, the first GRU layer in the trained generator model takes EEG features of dimension 30 as input and we take output from the TCN layer in the generator which produces EEG representations of dimension 32.  These EEG representations of dimension 32 are further used to perform continuous speech recognition experiments. 

The Figure 3 shows the generator and  discriminator model training loss.  The GAN model was trained for 201 epochs using adam optimizer with a batch size of 50.

\begin{figure}[h]
\begin{center}
\includegraphics[height=6.5cm, width=\linewidth,trim={0.1cm 0.1cm 0.1cm 0.1cm}]{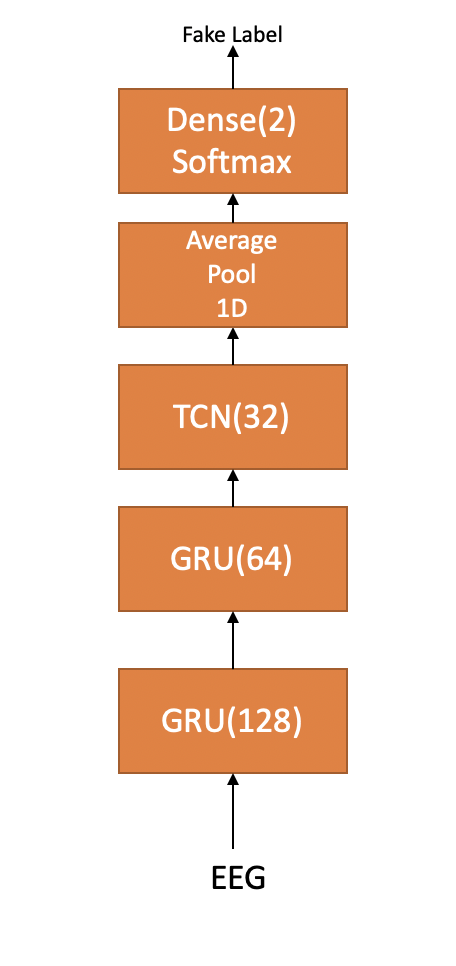}
\caption{Generator in GAN Model} 
\label{1vsall}
\end{center}
\end{figure}

\begin{figure}[h]
\begin{center}
\includegraphics[height=6.5cm, width=\linewidth,trim={0.1cm 0.1cm 0.1cm 0.1cm}]{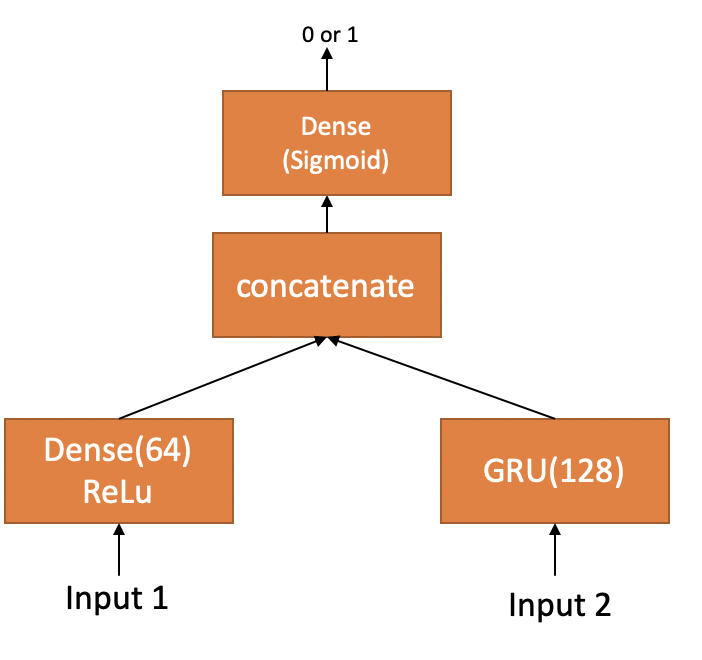}
\caption{Discriminator in GAN Model} 
\label{1vsall}
\end{center}
\end{figure}

\begin{figure}[h]
\begin{center}
\includegraphics[height=5cm, width=0.4
\textwidth,trim={0.1cm 0.1cm 0.1cm 0.1cm},clip]{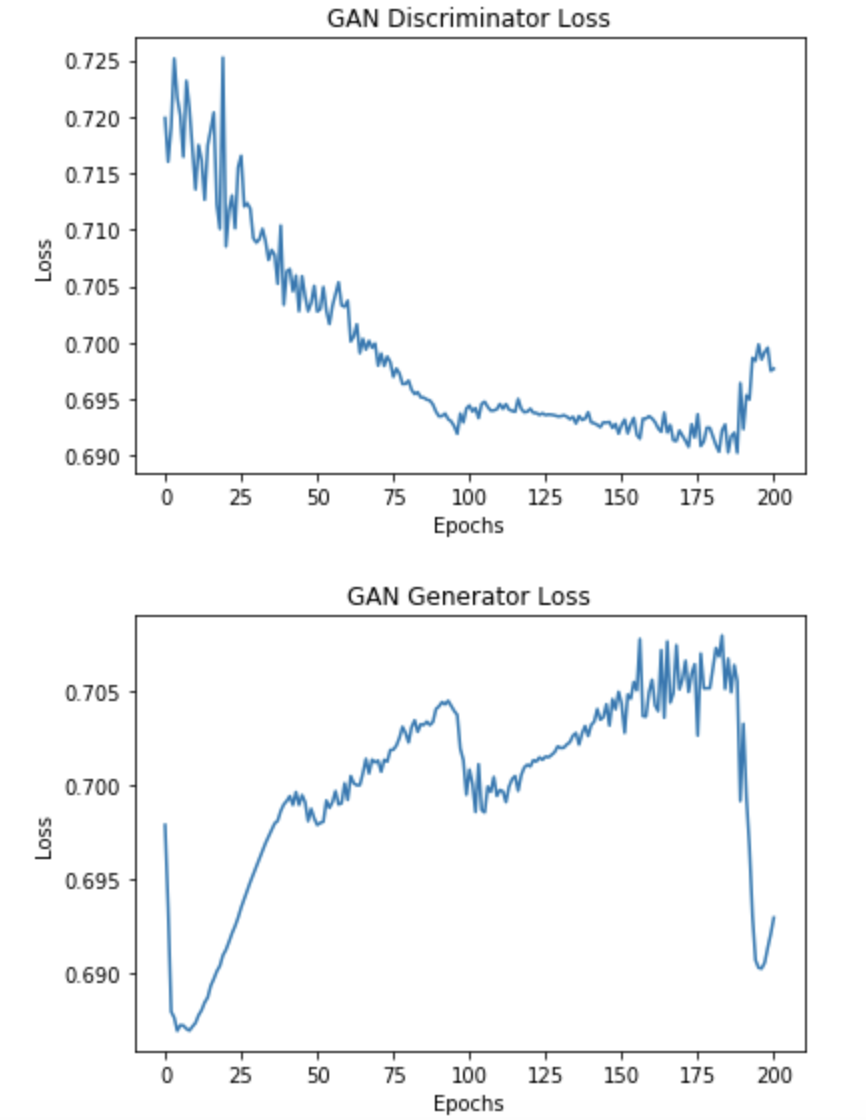}
\caption{GAN training loss} 
\label{1vsall}
\end{center}
\end{figure}

\section{ASR model used for performing continuous speech recognition experiments}
We performed continuous speech recognition using the raw EEG features of dimension 30 (baseline) and also using the EEG features of dimension 32 generated using the TCN layer of the generator described before. 

For performing experiments we used the connectionist temporal classification (CTC) \cite{graves2006connectionist,graves2014towards} model described in Figure 1 in \cite{krishna2019improving} with the exact same hyper parameters and training parameters used by authors in \cite{krishna2019improving} but the encoder layers in the CTC model were initialized with random weights \cite{krishna20,krishna2019state}. An external language model was used during inference time like the ones used by authors in \cite{krishna2019improving}.

\begin{figure}[h]
\begin{center}
\includegraphics[height=6.5cm, width=\linewidth,trim={0.1cm 0.1cm 0.1cm 0.1cm}]{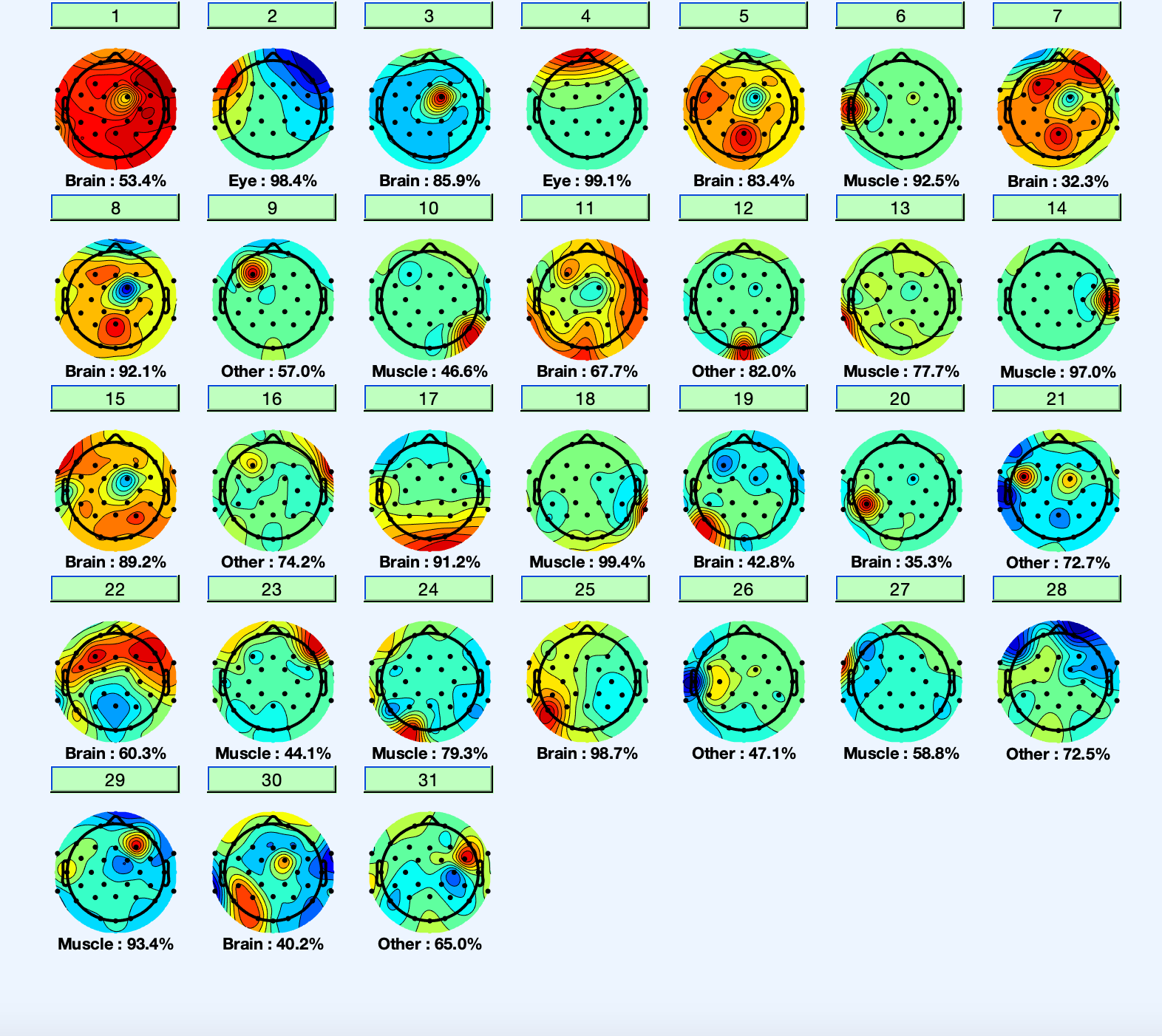}
\caption{ICA component classification showing various artifacts} 
\label{1vsall}
\end{center}
\end{figure}

\begin{figure}[h]
\begin{center}
\includegraphics[height=8.5cm, width=1\linewidth,trim={0.1cm 0.1cm 0.1cm 0.1cm}]{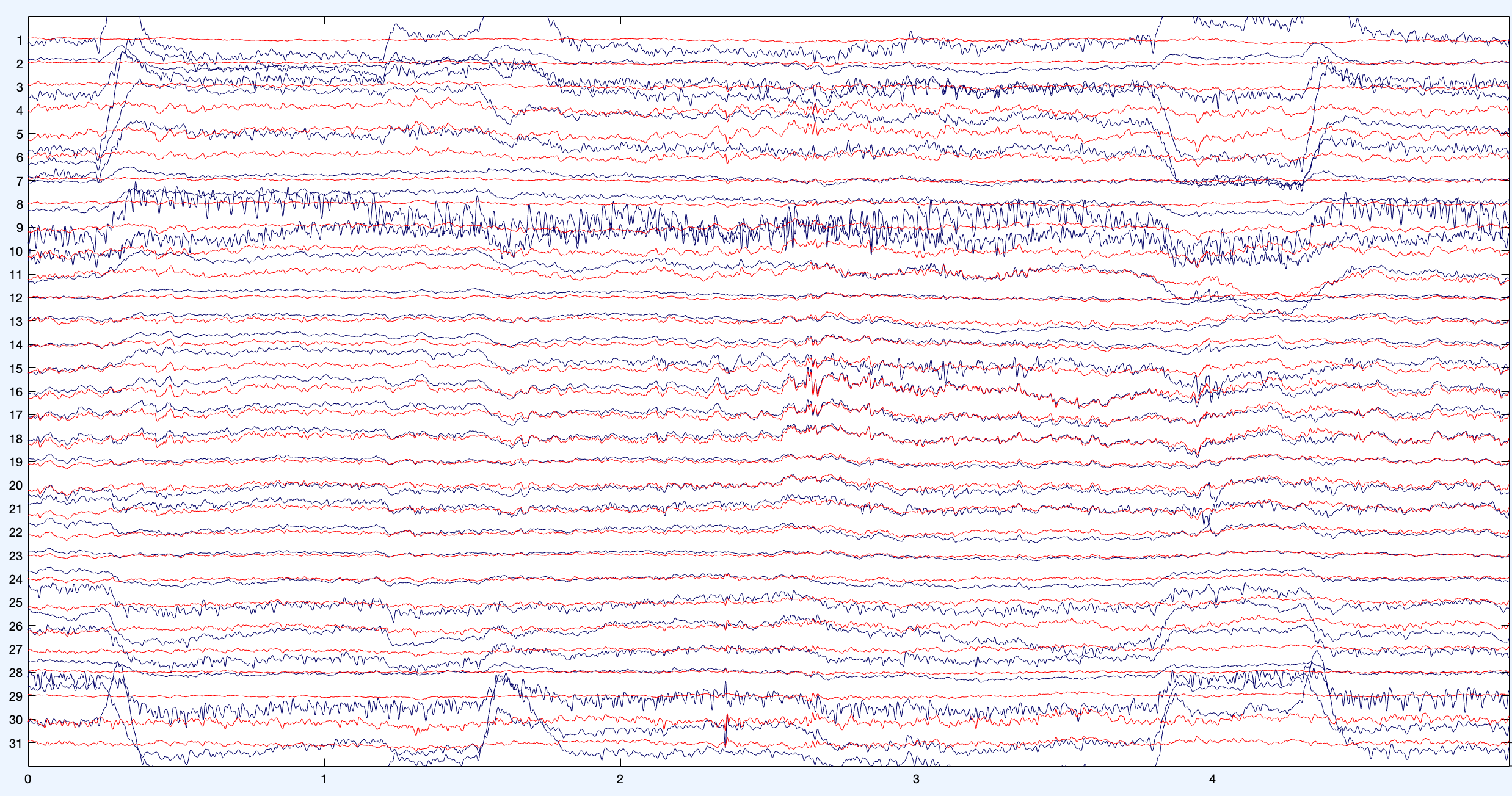}
\caption{EEG signals after removing artifacts. The blue color signal refers to raw signal before ICA artifact removal and \textbf{red color signal} refers to signal after ICA artifact removal} 
\label{1vsall}
\end{center}
\end{figure}


\section{Data Sets used for performing experiments}

For performing continuous speech recognition experiments using EEG, we used Data set A and B used by authors in \cite{krishna2019improving}. 
First we perform continuous speech recognition experiments using EEG features of dimension 30 from Data set A and B \cite{krishna2019improving} and then we pass the EEG features of dimension 30 to our GAN generator model described in Figure 1 and get the output from the TCN layer of the generator to get EEG features of dimension 32. Then experiments are performed using those EEG features of dimension 32.

For performing training the GAN model we used the combined EEG data for first two unique sentences from Data set A and B used by authors in \cite{krishna2019improving}, consisting of a total of 102 EEG recording examples. Since there were only two unique sentences, hence the generator model's final dense layer had two hidden units with softmax activation function. We considered EEG samples for only two unique sentences since we were interested in faster and stable training of the GAN model. 
More details of the data set, EEG experiment design, EEG recording hardware etc are covered in \cite{krishna20,krishna2019improving}. 

For each data set we used 80\% of the data as training set and remaining as test set. The train-test split was done randomly. There was no overlap between training and testing set. The way we splitted data for performing continuous speech recognition experiments in this work was exactly similar to the method used by authors in \cite{krishna2019improving}. 

\section{EEG feature extraction details}
We followed the same EEG preprocessing methods used by authors in \cite{krishna2019speech,krishna20} for extracting raw EEG features. 

The EEG signals were sampled at 1000Hz and a fourth order IIR band pass filter with cut off frequencies 0.1Hz and 70Hz was applied. A notch filter with cut off frequency 60 Hz was used to remove the power line noise.
The EEGlab's \cite{delorme2004eeglab} Independent component analysis (ICA) toolbox was used to remove other biological signal artifacts like electrocardiography (ECG), electromyography (EMG), electrooculography (EOG) etc from the EEG signals. The Figure 4 shows ICA component classification for various artifacts present in an EEG recording sample for a subject from Data set B. We can observe a significant presence of EMG artifact due to speech production. The Figure 5 shows the EEG signals after removing all artifacts for 31 channels. The figure shows a sample of 5 second recording from Data set B. 

We extracted five statistical features for EEG, namely root mean square, zero crossing rate,moving window average,kurtosis and power spectral entropy \cite{krishna2019speech,krishna20}. So in total we extracted 31(channels) X 5 or 155 features for EEG signals. The EEG features were extracted at a sampling frequency of 100Hz for each EEG channel.


\section{EEG Feature Dimension Reduction Algorithm Details}
After extracting EEG features as explained in the previous section, we used Kernel Principle Component Analysis (KPCA) \cite{mika1999kernel} to perform initial denoising of the EEG feature space as explained by authors in \cite{krishna20,krishna2019speech}. 
We reduced the 155 EEG features to a dimension of 30 by applying KPCA for both the data sets. We plotted cumulative explained variance versus number of components to identify the right feature dimension. We used KPCA with polynomial kernel of degree 3 \cite{krishna2019speech,krishna20}. We used these EEG features of dimension 30 as EEG features for calculating baseline results for continuous speech recognition experiments and then these 30 EEG dimensional EEG features are passed to the generator model described in Figure 1 during test time to get EEG features of dimension 32 from the TCN layer as output.

\section{Results}

We used word error rate (WER) as performance metric for continuous speech recognition experiments during test time. 

Table 1 shows the test time results obtained for continuous speech recognition experiments for Data set A. For baseline results we use 30 dimensional EEG features with CTC encoder with random weights, we then compare results obtained using our proposed method in this paper with the results obtained by authors in \cite{krishna2019improving}. We specifically compare our results with the results explained in Table 1 in reference \cite{krishna2019improving}. As seen from Table 1 continuous speech recognition using EEG features of dimension 32 generated using TCN layer in our GAN generator model described in Figure 1 always resulted in superior performance compared to baseline and demonstrated superior performance or lower WER compared to the method introduced by authors in \cite{krishna2019improving} for some of the test time experiments and for other cases our results were comparable with theirs. 

Table 2 shows the test time results obtained for continuous speech recognition experiments for Data set B. Similar observations seen in Table 1 were also noted for the results described in Table 2. 

Results from Tables 1 and 2 summarizes that our proposed method can be used to generate EEG features to improve the performance of continuous EEG based speech recognition systems. Our proposed method doesn't depend on additional features like acoustic or articulatory features like the method used by authors in \cite{krishna2019improving}.

\begin{table}[!ht]
\centering
\begin{tabular}{|l|l|l|l|}
\hline
\textbf{\begin{tabular}[c]{@{}l@{}}Total\\ Number\\ of\\ Sentences\end{tabular}} & \textbf{\begin{tabular}[c]{@{}l@{}}WER\\ (\%)\\ EEG\\ DIM\\ 30\\ BASE\\ LINE\end{tabular}} & \multicolumn{1}{c|}{\textbf{\begin{tabular}[c]{@{}c@{}}WER\\ (\%)\\ EEG\\ DIM\\ 30\\ REF\\ {[}1{]}\\ TECH\\ NIQUE\end{tabular}}} & \textbf{\begin{tabular}[c]{@{}l@{}}WER\\ (\%)\\ EEG\\ DIM\\ 32\\ PROPOSED\\ TECHNIQUE\end{tabular}} \\ \hline
21                                                                               & 82.93                                                                                      & 72.57                                                                                                                            & 75.75                                                                                               \\ \hline
42                                                                               & 77.66                                                                                      & 75.5                                                                                                                             & 75.71                                                                                               \\ \hline
63                                                                               & 85.78                                                                                      & 82.5                                                                                                                             & 83.19                                                                                               \\ \hline
84                                                                               & 86.3                                                                                       & 80.64                                                                                                                            & 74.72                                                                                               \\ \hline
105                                                                              & 97.05                                                                                      & 77.54                                                                                                                            & 80.88                                                                                               \\ \hline
126                                                                              & 103                                                                                        & 87.7                                                                                                                             & 85.01                                                                                               \\ \hline
\end{tabular}
\caption{Test time results for \textbf{Data set A}}
\end{table}

\begin{table}[!ht]
\centering
\begin{tabular}{|l|l|l|l|}
\hline
\textbf{\begin{tabular}[c]{@{}l@{}}Total\\ Number\\ of\\ Sentences\end{tabular}} & \textbf{\begin{tabular}[c]{@{}l@{}}WER\\ (\%)\\ EEG\\ DIM\\ 30\\ BASE\\ LINE\end{tabular}} & \textbf{\begin{tabular}[c]{@{}l@{}}WER\\ (\%)\\ EEG\\ DIM\\ 30\\ REF\\ {[}1{]}\\ TECH\\ NIQUE\end{tabular}} & \textbf{\begin{tabular}[c]{@{}l@{}}WER\\ (\%)\\ EEG\\ DIM\\ 32\\ PROPOSED\\ TECH\\ NIQUE\end{tabular}} \\ \hline
30                                                                               & 82.63                                                                                      & 74.36                                                                                                       & 76.58                                                                                                 \\ \hline
60                                                                               & 84.30                                                                                      & 74.45                                                                                                       & 74.24                                                                                                 \\ \hline
90                                                                               & 82.67                                                                                      & 77.76                                                                                                       & 77.83                                                                                                 \\ \hline
120                                                                              & 88.94                                                                                      & 79.68                                                                                                       & 79.00                                                                                       \\ \hline
150                                                                              & 90.39                                                                                      & 81.97                                                                                                       & 84.62                                                                                       \\ \hline
180                                                                              & 85.39                                                                                      & 84.9                                                                                                        & 84.35                                                                                        \\ \hline
\end{tabular}
\caption{Test time results for \textbf{Data set B}}
\end{table}

\section{Conclusion and Future work}

In this we paper we demonstrate that by making use of the ability of generative adversarial networks (GAN) to learn the loss function, the model can be trained to generate more meaningful EEG features from raw EEG features to improve the performance of EEG based continuous speech recognition systems. We compare our method with the method described by authors in \cite{krishna2019improving} to improve the performance of EEG based continuous speech recognition systems and we demonstrate that our method outperforms their method for some of the test time experiments and for other cases our results were comparable with theirs  when trained and tested using the same data sets and our proposed method doesn't need additional features like acoustic features or articulatory features which are needed to implement the method described by authors in \cite{krishna2019improving}. 

For future work we would like to improve the current results by adding CTC loss to our generator loss and also include a non differentiable external language model with the generator but that will require larger training data set with more number of EEG examples and data from larger number of subjects. We would also like to combine our proposed method with the method introduced by authors in \cite{krishna2019improving} to see if that helps in establishing a new baseline for state-of-the-art continuous EEG based speech recognition. 

\section{Acknowledgement} 
We would like to thank Kerry Loader and Rezwanul Kabir from Dell, Austin, TX for donating us the GPU to train the models used in this work.   

\bibliographystyle{IEEEtran}

\bibliography{mybib}


\end{document}